\documentclass[12pt]{article}
        \usepackage{amssymb}
        \usepackage{amsmath}
\def \d{\partial}
\def \bv{{\bf v}}

\def \brho{\boldsymbol{\rho}}
\def \bxi{\boldsymbol{\xi}}
\def \bomega{\boldsymbol{\omega}}

\begin{document}

\date{}

\author{ K.P. Zybin, V.A. Sirota, A.S. Ilyin, A.V. Gurevich }

\title{Generation of small-scale structures in the developed turbulence} \maketitle

\begin{abstract}
The Navier-Stokes equation for incompressible liquid is considered
in the limit of infinitely large Reynolds number. It is assumed
that the flow instability leads to generation of steady-state
large-scale pulsations.  The excitation and evolution of the
small-scale turbulence is investigated. It is shown that the
developed small-scale pulsations are intermittent. The maximal
amplitude of the vorticity fluctuations is reached along the
vortex filaments. Basing on the obtained solution, the pair
correlation function in the limit $r\to 0$ is calculated. It is
shown that the function obeys the Kolmogorov law $r^{2/3}$.
\end{abstract}

\section{Introduction}
In the turbulent flow, in addition to the average velocity of the
flow, random velocity pulsations are excited. These  pulsations
could be presented as a sum of different scales random movements.
The large-scale pulsations with scale $L$
of the same order as the characteristic parameters of the flow play a leading role
(For example, in the tube of radius  $R$ the scale is $L\sim
R/5$). Large-scale pulsations have the highest amplitudes.

The small-scale pulsations with scales $l<<L$ are excited also. They
have much smaller velocity amplitudes, and they could be
considered as a fine structure set to the main large-scale
movement. The small-scale pulsations contain only a small part of
the whole turbulent kinetic energy (see Landau, Lifchitz \cite{LL}, Monin,
Yaglom \cite{MY}).

If the viscosity of the liquid $\nu$ is small enough, and the
Reynolds number $R_u$ correspondingly is large, then the spectrum
of the small-scale pulsations becomes very wide. This type of
turbulence is  called developed. Let $\lambda_0$ be the
maximal scale where the viscosity is still significant; then
the range of scales $\lambda_0<<l<<L$ is called the inertial
interval. The  pulsations developed
inside the inertial interval in different scales are determined by nonlinear processes
only, since the viscosity $\nu$ is negligible. Therefore, it is possible
to study the inertial interval of the turbulence in the limit $\nu\to
0\,,\quad \lambda_0\to 0\,, \quad R_u\to \infty$ (Frisch
\cite{Fr}).

In the turbulent flow velocities are random. So, the
correlation functions could be used to describe them. 
Let us consider the isotropic turbulence. The pair correlation function
$$
K(r) = \left<[{\bf v}(\brho) - {\bf v}(\brho + {\bf r})]^2\right>,
$$
determines the relation between the values of velocity in two
near points  $\brho$ and $\brho+{\bf r}$. Since the turbulent
pulsations are isotropic, the correlation function  depends on the
distance $r$ between the points only.
 The pair correlation function measured in numerous experiments
has the universal form:
\begin{equation}\label{2/3a}
K(r) = C r^{2/3}
\end{equation}
The distance is restricted  by  the condition $r<<L$, i.e. the
experimental result (\ref{2/3a}) refers to the inertial interval
only. The Fourier-transform $S(k)$ of the correlation function 
(\ref{2/3a})  has been also investigated experimentally. These
investigations give
\begin{equation}\label{5/3}
S(k) = C_f k^{-5/3},
\end{equation}
where $k$ is the wave vector. The spectrum (\ref{5/3}) is called
the five-thirds law. The limit $r\to 0$ corresponds to
$k\to\infty$. In the developed turbulence the five-thirds law
is observed inside a  wide range of wave numbers -- up to
three-four orders of magnitude \cite{MY},\cite{Fr}.

We emphasize that the experimental measurements of the correlation
function in the small-scale region were initiated by the
theoretical predictions.
A.N. Kolmogorov in his fundamental works in 1941  \cite{Kol}
derived the expressions (\ref{2/3a}), (\ref{5/3}) for the velocity
correlation function in the homogeneous and isotropic turbulence.
\footnote{The law (\ref{5/3}), which is the direct consequence of
(\ref{2/3a}), was written in an explicit form in the papers by
A.M. Obukhov \cite{OB}.} The Kolmogorov's theory is
phenomenological. Its basic conception is the uniform dissipation
of energy in the turbulent liquid. There is a stationary flux of
energy in the Fourier space:  the energy is generated in
large-scale pulsations, and flows uniformly through the whole
inertial interval of scales. In this process, the flux of the energy is
conserved. The dissipation occurs only outside the inertial
interval, at the smallest "dissipative" scales $\le \lambda_0$.
Relaying on this physical model, using the relations of similarity 
and dimensions and the general properties of hydrodynamics equations,
the correlation function was found.

This fundamental  Kolmogorov's result was later confirmed in numerous
experiments. It stimulated a huge amount of theoretical,
mathematical and (in the recent time) numerical investigations. In
these works the theory of turbulence was widely developed 
(see the monographs \cite{MY},\cite{Fr},\cite{ZLF} -- \cite{Saff}
and literature therein). Recently, the methods of field theory and
solid-state physics have been used  \cite{L'v},\cite{Yah}.
However, the attempts to obtain the expression for the correlation
function directly from the Navier-Stokes equation without any
additional assumptions  have not been successful up to now (see
\cite{Fr} for more details).

Another approach to the problem is based on the
physical ideas of the leading role of singularity in the
small-scale structures of developed turbulence \cite{Fr},
\cite{Kuz}. However, despite significant efforts in this
direction, neither  the correlation function has been derived from
the Navier-Stokes equation, nor even the existence of singular
solutions has been proved.

Thus, the problem of derivation of the fundamental Kolmogorov's
result directly from hydrodynamic equations is not solved yet (see
the monographs \cite{MY},\cite{Fr}).

In this paper we propose a new approach to the problem. It allows
to find the structure of the small-scale turbulence and the pair
correlation function. 

From the hydrodynamic equations written in the Lagrangian
reference frame, we derive the equation describing the joint
probability density of vorticity $\bomega=\nabla \times {\bf v}$
and its time derivative. We show that moments of the vorticity
distribution grow unrestrictedly in time. Then we find an
asymptotic solution at infinitely large time. Basing on it, we obtain the
spatial distribution of the vorticity  where it is
large. These are  vortex filaments. They  give the main
contribution to the pair correlation function.

The paper is organized as  follows.

In Section 2 the equations of motion of incompressible liquid are
considered. Their decomposition  in the vicinity of trajectory of
an arbitrary lagrangian particle is written. It is shown that
{\it local vorticity growth} is determined by {\it anisotropic part of large-scale
pulsations} of pressure.

In Section 3, supposing the randomness of large-scale
pulsations of pressure,  the equation for probability density of
vorticity and its time derivative is obtained. We show that even
moments of the vorticity distribution grow exponentially, the
higher moments growing faster than the lower ones. This is 
the manifestation of intermittency of hydrodynamic turbulence in small scales.

In Section 4 we find the large time asymptotic  solution for
the joint probability density of  vorticity and its time
derivative.

In Section 5, on the ground of the obtained asymptotic solution,
the spatial structures contributing  mainly to the
asymptote of the probability density are investigated.
We show that these are the vortex filament structures which
determine the pair correlation function of turbulent pulsations
(\ref{2/3a})  as $r\to 0$.

In  Conclusion we formulate and discuss the main results of the paper.

\section{The  statement of the problem}

Let us consider the Navier-Stokes equation for incompressible
liquid. It is known that at the scales larger than the viscous
scale   $\lambda_0$ it takes the form of the Euler equation:
\begin{equation} \label{hd}
\frac{\d \bv }{\d t} + (\bv \cdot \nabla ) {\bv} + \frac{\nabla
p}{\rho} = 0 \,; \qquad \nabla \cdot \bv =0
\end{equation}
Here  $\bv$ is the velocity of the flow, $p$ is the  pressure. The
density $\rho$ is taken unity below. The second 
equation expresses the incompressibility of the liquid. The
equations (\ref{hd})  describe the processes on the scales inside
the inertial interval (see \cite{Fr}). From
(\ref{hd}) one can find the relation connecting the pressure with
the flow velocity:
%\begin{equation} \label{pressure}
$$
-\Delta p = \nabla _i v _j \cdot \nabla _j v _i
$$
%\end{equation}

To investigate the local properties of the turbulent flow we pass
on to a coordinate system co-moving to some element of the liquid
with coordinates  $\bxi(t)$:
$$
{\bf r'}={\bf r}-  \bxi (t) \,, \quad {\bf v'}={\bf v} - \dot \bxi
; \quad \ddot \bxi = - \nabla p \left| \begin{array}{l} \\
{\bf r} = \bxi (t) \end{array} \right.
$$
Dot  means the time derivative. After such change of
variables the equations (\ref{hd}) take the form
\begin{equation} \label{hdmoving}
\frac{\d \bv' }{\d t} + (\bv' \cdot \nabla ) \bv' + \nabla P = 0 \,,
 \qquad \nabla P = \nabla p + \ddot  \bxi \,, \qquad \nabla
\cdot \bv' =0
\end{equation}
Since the reference frame is chosen to be co-moving, at the point
${\bf r}=\bxi (t)$ we have
$$
\nabla P ({\bf r'=0}) =0 \,, \qquad {\bv'} ({\bf r'=0})=0
$$
Expanding the velocity  $\bv'$ and the pressure $P$  into  a Taylor
series in the vicinity of the co-moving point we have the main term:
\begin{equation} \label{Taylor}
v'_i= \left( \frac 12 \varepsilon_{ikj} x_k  +b_{ij} \right) r'^j
\end{equation}
\begin{equation} \label{TaylorP}
P={1\over 2} \rho_{ik} r'^i r'^k \,, \qquad \rho_{ik}=\nabla_i
\nabla_k P
\end{equation}
Here the tensor $\left.\frac{\partial v'_i}{\partial
r'^j}\right|_{r'=0}$ is decomposed into a sum of symmetric
$b_{ij}$ and antisymmetric $\frac 12 \varepsilon_{ikj} x_k$ parts.
Note that $b_{ii}=0$ since  $\nabla \bv' =0$; $\rho_{ij}$ is
symmetric. It is easy to check that the vector $x_i$ defined by
the asymmetric part of  $\partial v'_i /
\partial r'^j$ is the vorticity  ($\bomega = \nabla \times {\bf
v}$) of the flux at the point  $\bxi$:
$$
\omega_i \left| \begin{array}{l} \\  {\bf r' } =0 \end{array}
\right. = x_i  .
$$

Combining (\ref{Taylor}),(\ref{TaylorP}) and (\ref{hdmoving}), we
obtain
$$
\left[ \left( \dot b_{ij} + \frac 14 (x_i x_j - x^2) \delta_{ij}
+b_{ik} b_{kj} + \rho_{ij} \right) + \frac 12 \left(
\varepsilon_{ikj} \dot x_k - \varepsilon_{jkn} x_k b_{in} +
\varepsilon_{ikn} x_k b_{jn} \right) \right] r'^j =0
$$
The first term in the square brackets is symmetric, the second
term is antisymmetric.  Since the equation should hold for any
${\bf r'}$, both terms are equal to zero.  Actually, multiplying
the expression in the brackets by   $\epsilon_{ijn}$ and taking
 $b_{ii}=0$ into account, we find for $x_i$ and $b_{ij}$:
\begin{equation} \label{b}
\dot b_{ij} + \frac 14 (x_i x_j - x^2 \delta_{ij}) +b_{ik} b_{kj}
+ \rho_{ij} =0
 \end{equation}
\begin{equation} \label{a}
\dot x_n  = b_{nk} x_k
 \end{equation}
Taking the time derivative of (\ref{a}), we obtain finally the
equations set for the three components $x_i$:
\begin{equation}  \label{omega}
\ddot x_n = - \rho_{nk} x_k
\end{equation}

Let us clarify now  the physical meaning of the symmetric part of
velocity $b_{ik}$. Namely, let us express it in terms of  space
distribution of the vorticity $\bomega ({\bf r})$. We shall see
that in the completely isotropic flow  $b_{ij} = 0$. Hence in
accordance with   (\ref{a}), the vorticity at the lagrangian
point does not change. In the
real flow there are two regions where the isotropy may be broken:
either local, on account of small-scale pulsations of pressure in
the vicinity of the point under consideration; or global, the
remote areas close to the boundary  of the system, at the scales
of  the order $R$. We shall show that $b_{ij}$  is determined just
by the global break of isotropy.

%─ы  ¤Єюую,
% ъръ ёыхфєхЄ шч (\ref{Taylor}), эрь эхюсїюфшью т√ўшёышЄ№ ЄхэчюЁ
%$\left.\frac{\partial v_i}{\partial r_k}\right|_{r=0}$.
Since $\nabla \cdot {\bf v} =0$, there exists a vector potential ${\bf A}$:
$$
{\bf v} =
\nabla \times \, {\bf A}\,,\quad \nabla \cdot \, {\bf A} = 0.
$$
Then
\begin{equation}\label{9a}
\Delta {\bf A} = -\bomega.
\end{equation}
To separate the singularity accurately, let us expand ${\bf A}({\bf
r})$ and ${\bomega}({\bf r})$ into a series on spherical harmonics:
$$
{\bf A} = \sum \limits _{l=1}^{\infty} \sum \limits_{m=-l}^{m=l}
{\bf A}_{lm} (r) Y_{lm}(\theta, \varphi) \,, \qquad {\bomega} =
\sum \limits _{l=1}^{\infty} \sum \limits_{m=-l}^{m=l} {\bomega
}_{lm} (r) Y_{lm}(\theta, \varphi)  .
$$
The solution of the Poisson equation (\ref{9a}) is:
\begin{equation}\label{davl1}
{\bf A}_{lm}(r)= \frac{ r^{-l-1}}{2l+1} \int _0^r \bomega_{lm}(r_1)
r_1^{l+2}dr_1 + \frac{r^{l}}{2l+1}\int_r^{\infty} \bomega_{lm}(r_1)
r_1^{1-l} dr_1
\end{equation}
The integration limits are chosen  to provide the
convergence of the integrals at $r\to 0$ and $r\to \infty$. Note
that
%шэЄхуЁры (\ref{davl1}) эх шьххЄ ЁрёїюфшьюёЄш яЁш $r\to 0$, Є.ъ.
for analytic function $\bomega_{lm}(r) \propto r^l$ as $r\to 0$.

To evaluate $b_{ij}$, we need to determine the limit   $ \nabla_k v_i
 =  \varepsilon_{ijn} \nabla_k \nabla_j A_n $ as
$r\to 0$. Only quadratic (in the coordinates $r_i$) part of $\bf
A$ contributes in it. This quadratic part consists of two terms
proportional to the zeroth  ${\bf A}_{00}$ and the second ${\bf
A}_{2m}$ spherical harmonics. Hence, we are interested in the 
two harmonics only.

The zeroth harmonic ${\bf A}_{00}$ gives the local contribution
corresponding to the antisymmetric part of the velocity tensor:
${\bf x}=\left. \bomega_{00}\right| _{r=0} = -\left. \Delta {\bf
A}_{00} \right| _{r=0}$. However, in the symmetric tensor $b_{ij}$
its quadratic component ${\bf A}_{00}\propto r^2$ is cancelled: it
is just the fact that gives $b_{ij}= 0$ in isotropic medium. So,
only the second harmonic  ${\bf A}_{2m}$ remains. Since
$\bomega_{2m}(r) \sim r^2$ as $r\to 0$, we see that the first
integral in (\ref{davl1}) behaves like $r^4$, and the second one -
like $r^2$ as $r\to 0$. Hence, the contribution of the "local"
item is negligibly small, and the symmetric part of the velocity
tensor  $b_{ij}$ is determined by "global", large-scale properties
of the whole flow.
%яЁюяюЁЎшюэры№эр ътрфЁєяюы№эющ ёюёЄрты ■∙хщ яюЄхэЎшрыр:
 \footnote{Note that in two-dimensional flow such a division into local and large-scale
 components is not possible. The zeroth cylindrical harmonic   ${\bf
A}_{0}$ diverges logarithmically, and as a result the "local"
component should influence on the large-scale component. }
Returning to the rectangular coordinates and taking the
derivative, we obtain
$$
b_{ij}=  \varepsilon_{jnk}  \int\frac{ \omega_n ({\bf r'})}{r'^3}
\left(\delta_{ik}-3\frac{r_i'r_k'}{r'^2}\right) d{\bf r'} + \left( i \leftrightarrow j\right)
$$
According to our analysis, the integrand has no singularity at
$r=0$; the integral accumulates at the scales of the order of $R$,
where  the isotropy breaks.

The analogous argumentation shows that the pulsations of pressure
$\rho_{ik}$ (see (\ref{b}))  could also be presented as a sum of
local and large-scale pulsations; the local part of the tensor is
$x_i x_k-\delta_{ik}x^2$. From (\ref{omega}) it follows that this
tensor does not affect the local vorticity of the flow. Hence, the
local dynamics of vorticity (\ref{omega}) depends on the
large-scale pulsations of the  pressure $\rho_{nk}$ only.

Thus we obtain the first main property of the turbulent flow:
{\it the local vorticity} along the streamline in homogeneous and isotropic
flow is {\it determined by anisotropic part of large-scale pulsations
of the pressure}.

\section{Probability density equation}

Since we interested in  statistical properties of the flow, let
us introduce the probabilistic description. We consider now  the
vorticity $\bomega(t)$ as a random quantity.  Its change still obeys to
(\ref{omega}). Instead of one equation of the second order, let us
consider a system of two first-order equations:
\begin{equation}\label{xy}
    \dot{x_i}=y_i \, \qquad \dot{y_i}=-\rho_{ij} x_j
\end{equation}
Here $x_i \equiv\omega_i$ and $y_i \equiv\dot{\omega}_i$. We
introduce a joint probability density
\begin{equation}   \label{sovmestxy}
f(t,{\bf x},{\bf y}) = <\delta({\bf x}-{\bf x}(t))\delta({\bf
y}-{\bf y}(t)))>  .
\end{equation}
Here ${\bf x}(t), {\bf y}(t)$ are the solutions of (\ref{xy}) at
the given realization of $\rho_{ij}$ and initial conditions; the
average is taken over the ensemble of all possible realizations.

The aim of the paper is to study a steady-state
turbulent flow in the inertial interval of scales, i.e.  at scales
$l$ and time $t$ satisfying to the conditions
\begin{equation} \label{uslovia}
l<<L,\qquad\quad t>>\tau_c
\end{equation}
Here $L$ and $\tau_c$ are the characteristic space and time
correlation scales of the large-scale vortices.  These large-scale
vortices depend on the specific geometry of the installation and
on the boundary conditions. According to the experimental data
 \cite{NWL}, the large-scale velocity pulsations are random  and Gaussian.
Thus, in the equations (\ref{omega}) or (\ref{xy}) for the local
vorticity, the matrix $\rho_{ij}(t)$ describing the large-scale fluctuations
of pressure could be taken Gaussian and, because of
(\ref{uslovia}),  delta-correlated in time. These propositions would be discussed
below.

Note that, as it follows from (\ref{omega}),  the "random"
behavior of vorticity (or velocity) is caused by the
randomness of the large-scale flow and the corresponding matrix
$\rho_{ik}(t)$.

The Gaussian random process is described by a pair correlation
function

\begin{equation}\label{2a}
<\rho_{ij}(t) \rho_{kl}(t')> = D_{ijkl}\,\delta(t-t')
\end{equation}
Using (\ref{xy}) and taking time derivative of the probability
density function, we obtain
\begin{equation}\label{3a}
 \frac{\d f}{\d t} +  y_k \frac{ \d f}{\d {x_k}}=x_p\frac{\partial}{\partial y_k}\left<
 \rho_{kp} \delta({\bf x}-{\bf x}(t))\delta({\bf y}-{\bf y}(t))\right>
\end{equation}
Let $R({\bf x},{\bf y},\rho)$ be a functional of  $\rho$.  To find
the correlation function $\left<\rho_{kp}R({\bf x},{\bf
y},\rho)\right>$ we use the standard averaging technics for
delta-correlated random process (see the monograph by 
Klyackin, \cite{KL}):
$$
\left<z_k R[z]\right> = \sum_{k'}\int dt'
\left<z_k(t)z_{k'}(t')\right> \left<\frac{\delta R[z,t]}{\delta
z_{k'}(t')}\right>
$$
Taking (\ref{2a}) into account, we get:
\begin{equation}\label{5a}
\left<\rho_{kp}R({\bf x},{\bf y},\rho)\right>=\sum_{k'p'}D_{k p
k'p'} \left<\frac{\delta R[{\bf x},{\bf y},t]}{\delta
\rho_{k'p'}(t)}\right>
\end{equation}
To evaluate the variational derivative (\ref{5a}), we use the
equations of motion (\ref{xy}); it follows
$$
\left.\frac{\delta y_k(t)}{\delta\rho_{k'p'}(t')}\right|_{t=t'}
=-\delta_{kk'}x_{p'}(t)\,,\qquad \left.\frac{\delta
x_k(t)}{\delta\rho_{k'p'}(t')}\right|_{t=t'}=0
$$
Combining this with (\ref{3a}), we obtain the Fokker-Planck
equation for the function $f(t,{\bf x},{\bf y})$:
\begin{equation}\label{ijkl}
 \frac{\d f}{\d t} + { y_k} \frac{ \d f}{\d { x_k}}=D_{ijkl} x_j x_l\frac{\partial^2 f}{\partial
 y_i\partial y_k}
\end{equation}
The matrix $\rho_{ij}(t)$ is symmetric (\ref{TaylorP}). Hence, in
the homogeneous and isotropic medium the general form of the
matrix  $D_{ijkl}$ is
\begin{equation}\label{<ik>}
D_{ijkl}= D \delta (t-t') \left( \delta_{ik}\delta_{jl}
+\delta_{il}\delta_{jk} + \Gamma \delta_{ij}\delta_{kl} \right)
\end{equation}
The constants $D$ and $\Gamma$ depend on the large-scale flow. In
addition to the isotropy and homogeneity, it is natural to suppose
statistical independence of different  components  $\rho_{ik}$. In
this case one has $\Gamma=0$. However, the values $D$ and $\Gamma$
appear to be unimportant. The parameter  $D$ in the equation
vanishes as a result of time normalization.  As we shall see
below, the resulting properties of the turbulence depend only
weakly on the parameter  $\Gamma$ (the  positive
definiteness  leads to a restriction
$\Gamma> -2$).

Substituting (\ref{<ik>})  into (\ref{ijkl}),  we obtain finally:
\begin{equation}\label{f}
 \frac{\d f}{\d t} + { y_k} \frac{ \d f}{\d { x_k}} = \left[x^2\frac{ \d^2 f}{ \d {\bf y}^2 }+
    \gamma \left( { x_k} \frac{\d}{\d y_k} \right)^2 f\right]
\end{equation}
Here $\gamma=1+\Gamma$, time $t$ is normalized by  $D^{1/3}$. The
value $D^{-1/3}$ is the characteristic time  of
probability density change. As it was shown above, in the completely
isotropic turbulence $D=0$. Taking into account small anisotropy,
we have $D^{-1/3} >> \tau_c$. This allows to use the delta-correlation 
approximation in derivation of the equation (\ref{f}).

We now itemize the main properties of the equation (\ref{f}):

1. All momenta of values $x_k$ and $y_j$ of the order $n$ are
connected  by a system of the first order linear differential equations.

2. Even momenta grow exponentially. Independently of the initial
conditions, the function $f$ at large values $t$ depends on the
modules $x,\, y$ and the cosine of the angle between the vectors
$\mu=({\bf x},{\bf y})/xy$ only.

3. The higher even momenta grow faster than the lower ones.

To illustrate these statements, consider the momenta of the second
and the fourth order. Integrating the equation (\ref{f}) in ${\bf
x}$ and in ${\bf y}$, we obtain for the second-order momenta:
$$
\frac{d}{dt}<x_ix_j>=<x_iy_j + x_jy_i>
$$
\begin{equation}\label{<xy>}
\frac{d}{dt}<x_iy_j + x_jy_i>=2<y_iy_j>
\end{equation}
$$
\frac{d}{dt}<y_iy_j>=2\delta_{ij}<x^2> + 2\gamma<x_ix_j>
$$
Let us consider the invariant momenta of the second order, i.e.
$<x^2>$ , $<y^2>$  and $<{\bf x}\cdot {\bf y}>$. Their evolution
is determined by characteristic equation
$$
\lambda^3 -4\Gamma -16 = 0.
$$
Asymptotically as  $t\to\infty$ we obtain
$$
<x^2>\propto<y^2>\propto<{\bf x}\cdot{\bf y}>\propto\exp
(\Lambda_2 t)
$$
Here  $\Lambda_2=(16+4\Gamma)^{1/3}$.

For the other momenta ($i\ne j$) one has
$$
\lambda_2^3 = 4 + 4\Gamma
$$
We see that $\Lambda_2>\lambda_2$. Hence, at large time the
invariant momenta are much larger than the others. In other words,
the  probability density function at large time depends on three
variables $x, y, \mu=x_i y_i/(x y)$ only.

The characteristic equation for invariant momenta of the fourth
order takes the form:
$$
\lambda^6-(84\Gamma + 244)\lambda^3 -1280 = 0.
$$
For example, for $<x^4>$ one has $<x^4>\propto \exp(\Lambda_4 t)$,
where
$$
\Lambda_4 = \frac{1}{8}\left[ \left((84\Gamma + 224)^2 + 5120
\right)^{1/2} + (84\Gamma + 224)\right]^{1/3}
$$
One can check up that for the values  $\Gamma
>-0.9 $ holds $\Lambda_4>2\Lambda_2$. Hence, as
$t\to\infty$ one has $ <x^4> \gg <x^2>^2$.

The obtained relations demonstrate that the higher momenta of the
vorticity module grow exponentially in time. This property is
called intermittency. It reveals itself in the instability of
small-scale flow. Physically this instability means that under the
influence of large-scale random pulsations a drop of
incompressible liquid stretches out. This leads to generation of
vortex filaments. These filaments provide the basis of the small-scale
turbulence (a simple physical example demonstrating the process of
filament growing is considered in  Appendix).

It will be shown later that the domain of  parameters  $y \gg x$ 
plays an especially important role. In this domain
the suggestion of gaussian random process is
not needed.  Actually, from (\ref{xy}) it follows that the change
of  $y$ during the correlation time  is  $\Delta y \sim x$. Hence,
\begin{equation}\label{dy}
\frac{\Delta y}{y}\approx \frac{x}{y} \ll 1 \qquad
\hbox{if}\,\quad y \gg  x
\end{equation}
In this case the fluctuations of the probability function are very
small: $\delta f \ll <f>$, and one can obtain the equation
(\ref{f}) using the perturbation theory. As a result, the equation
 has the Fokker-Planck form.

\section{Asymptotic form of the probability density function}

As it was shown in the previous section, the probability density
function (\ref{f}) at large time  depends on three variables
only:
 $f({\bf x},{\bf y})=f(x,y,\mu)$, where $\mu=({\bf x},{\bf y})/x y$.
 Besides, the equation (\ref{f}) and the initial conditions to the
 probability function allows integrating over  three other
 variables.

As a result, the equation (\ref{f}) takes the form:
\begin{eqnarray}\label{xym}
\frac{\d f}{\d t} + \frac y{x^2} \frac{\d}{\d x} \left( \mu x^2 f
\right) + \frac yx \frac{\d}{\d \mu} \left( (1-\mu^2) f \right) =
\frac {x^2}{y^2} \frac{\d}{\d y} \left( y^2 \frac{\d f}{\d y}
\right) + \frac {x^2}{y^2} \frac{\d}{\d \mu} \left( (1-\mu^2)
\frac{\d f}{\d \mu} \right) \\
\nonumber + \gamma \left( \mu x \frac{\d}{\d y} + \frac{x}{y}
(1-\mu^2) \frac{\d}{\d \mu} \right)^2 f
\end{eqnarray}

The function $f$ must satisfy the normalization condition
 $\int f d{\bf x} d{\bf y} = \int f x^2
y^2 dx dy d\mu =1$, and two conditions of zero flux from the
boundaries $x=0$ and $y=0$. Let us specify the meaning of these
conditions. For that  we return to the equation
(\ref{f}). It has the divergent form $\partial f/\partial
t=\nabla_{\alpha} {\bf J}^{\alpha} \,,  \alpha=1..6$.  The flux
density ${\bf J}^{\alpha}$ in a 6-dimensional space  $({\bf x},{\bf
y})$ is
$$
{\bf J} =\left\{ -{\bf y}f , x^2 \frac{\d f}{\d \bf y}  + \gamma
{\bf x} \left( {\bf x} \cdot \frac{\d f}{\d \bf y}\right) \right\}
$$
The no-flow boundary condition at $y=0$ means that the integral of
$\bf J$ over the 5-dimensional surface  ${|{\bf y}|=\epsilon}$
vanishes as $\epsilon \to 0$. After integrating in all angles
$d\Omega_x d\Omega_y=4\pi \cdot 2\pi d\mu$, in terms of variables
$x,y,\mu$ we have
\begin{equation} \label{granicay=0}
\int \left( x^2 \frac{\d f}{\d \bf y}  + \gamma {\bf x} \left(
{\bf x} \cdot \frac{\d f}{\d \bf y}\right)\right) \cdot \frac{\bf
y}{y} x^2 dx d\Omega_x y^2 d\Omega_y = 8\pi^2 \int  (1+\gamma
\mu^2) \frac{\partial f}{\partial y} x^2 y^2 dx d\mu
\begin{array}{c}
\\ \longrightarrow \\ y\to 0 \end{array} 0
\end{equation}
Similarly, the no-flow condition at  $x=0$ leads to
\begin{equation} \label{granicax=0}
\int \mu y f x^2 y^2 d \mu dy
\begin{array}{c} \\ \longrightarrow \\ x\to 0 \end{array} 0
\end{equation}
The expressions (\ref{granicay=0}) and (\ref{granicax=0}) are the
boundary conditions for the equation (\ref{xym}).

We now return to (\ref{xym}). We search for a stationary solution
as $t\to \infty$. Choosing a new variables $x,\,
z=y^3/(3x^3), \, \mu$ let us present the function $f(x,y,\mu)$
in the form
\begin{equation}\label{fF}
f(x,y,\mu)= \sum \limits_{\alpha} x^{-2} x^{-\alpha} F(z,\mu;
\alpha)
\end{equation}
The set of eigenvalues $\alpha$ is to be found by solution of (\ref{xym}) with
boundary conditions (\ref{granicay=0}), (\ref{granicax=0}).

The equation (\ref{xym}) then takes the form
\begin{equation}\label{F}
z\frac{\partial^2F}{\partial z^2} +\left(\frac{4}{3} + \mu
z\right) \frac{\partial F}{\partial z} +\frac{\alpha}{3}\mu F
-\frac{1}{3}\frac{\partial}{\partial \mu}\left[(1-\mu^2)F\right]
+\frac{1}{9
z}\frac{\partial}{\partial\mu}\left[(1-\mu^2)\frac{\partial
F}{\partial\mu}\right]
\end{equation}
$$
+\gamma \left[  z \left( \mu \frac{\d}{\d z} + \frac{1-\mu^2}{3z}
\frac{\d}{\d \mu} \right)^2 F + \frac 23 \mu \left( \mu
\frac{\d}{\d z} + \frac{1-\mu^2}{3z} \frac{\d}{\d \mu} \right) F
\right] =0
$$
Integrate  (\ref{F}) on variable $\mu$ and define
functions $\overline{\mu}(z)$ ш $\overline{\mu^2}(z)$:
\begin{equation}\label{opr}
\overline\mu(z)=\frac{\int_{-1}^{1} \mu F\,d\mu}{\int_{-1}^{1}
F\,d\mu}= \frac{F_1}{3 F_0} \, \quad
\overline{\mu^2}(z)=\frac{\int_{-1}^{1} \mu^2 F\,d\mu}{\int_{-1}^{1}
F\,d\mu}= \frac{2 F_2}{15 F_0} + \frac 13
\end{equation}
Here $F_k (z)$ are coefficients in Legendre expansion of the
function $F$. The equation (\ref{F}) takes the form
\begin{equation}\label{Fmu}
z (1+\gamma \overline{\mu^2}) F_{0zz} + \left(\frac43  + {\overline \mu}z
+\gamma \left( 2z \overline{\mu^2}_z + \frac{7 \overline{\mu^2}-1}3 \right)
\right) F_{0z}
\end{equation}
$$
+ \left(\frac{\alpha}{3}\overline{\mu} + z{\overline\mu_z} +\gamma \left( z
\overline{\mu^2}_{zz} +\frac 73 \overline{\mu^2}_z  + \frac{3 \overline{\mu^2}
-1}{9z} \right) \right) F_0 = 0 
$$

Substituting $F_0(z)=w(z)\,exp\left(-\int_0^z
\frac{\overline\mu(p)\,dp}{1+\gamma \overline{\mu^2}}\right)$, we get
\begin{equation}\label{w}
(1+\gamma \overline{\mu^2}) zw_{zz} + \left(\frac43 +\gamma
\frac{7\overline{\mu^2}-1}{3} + 2\gamma z \overline{\mu^2}_z -{\overline\mu}
z\right) w_z + \left(\frac{\alpha}{3} - \frac{4}{3(1+\gamma
\overline{\mu^2})} \right) {\overline\mu} w
\end{equation}
$$
 + \gamma \left( z\overline{\mu^2}_{zz} + \frac 73 \overline{\mu^2}_z +
\frac{3\overline{\mu^2}-1}{9z} -\frac{z\overline{\mu}
\overline{\mu^2}_z}{1+\gamma \overline{\mu^2}} - \frac{\overline{\mu}
(7\overline{\mu^2}-1)}{3(1+\gamma \overline{\mu^2})} \right) w = 0
$$
The solutions of (\ref{w}) could be presented as a series
 \footnote { Zero is a regular critical point of (\ref{w}), since
$\overline{\mu}(z)$ and  $\overline{\mu^2}(z)$ are unbounded, and
$1+\gamma>0$.}
 $w=z^s \sum \limits_{n=0}^{\infty} c_n z^n$.
It converges on the domain $0<z<\infty$ (if $\overline{\mu}(z)$ and
$\overline{\mu^2}(z)$  have no singularity).

In order to find $s$ and $c_n$, let us consider the asymptote $z
\to \infty$ \footnote{As it has been mentioned in the end of
previous section, in this limit the equation for probability
function has the Fokker-Planck form independently of statistical
properties of the large-scale random process}. Expanding
(\ref{Fmu}) into Legendre series, we get
$$
z F_{0zz} + \frac43 F_{0z} +\frac13\left[ z F_{1z} +
\frac{\alpha}{3} F_1\right] = 0 \,
$$
$$
z F_{1zz} + \frac43 F_{1z} +\left(z\frac{\partial}{\partial z} +
\frac{\alpha}{3}\right) \left(F_0 + \frac25 F_2\right) + \frac23
F_0 -\frac{2}{15} F_2 = \frac{2}{9 z}F_1 \,
$$
$$
..........
$$
$$
zF_m''+\frac 43 F_m' + \left( z\frac{\partial}{\partial z} +
\frac{\alpha}3 \right) \left( \frac {\scriptstyle m}{\scriptstyle 2m-1} F_{m-1} +
\frac{\scriptstyle m+1}{\scriptstyle 2m+3} F_{m+1} \right)
$$
$$
+ \frac 13 \left( \frac {\scriptstyle m(m+1)}{\scriptstyle 2m-1} F_{m-1} -
\frac{\scriptstyle m(m+1)}{\scriptstyle 2m+3} F_{m+1} \right) -
 \frac{\scriptstyle m(m+1)}{\scriptstyle 9z} F_m
$$
$$
+ \frac {\gamma}{9z} \left(
\frac{\scriptstyle m^2(m-1)(m-2)}{\scriptstyle (2m-1)(2m-3)}F_{m-2} - m(m+1)
\frac{\scriptstyle 2m(m+1)-1}{\scriptstyle (2m-1)(2m+3)}F_m
+\frac{\scriptstyle (m+1)^2(m+2)(m+3)}{\scriptstyle (2m+3)(2m+5)} F_{m+2} \right)
$$
$$ + \gamma z  \left(
\frac{\scriptstyle m(m-1)}{\scriptstyle (2m-1)(2m-3)}F''_{m-2} +
\frac{\scriptstyle 2m(m+1)-1}{\scriptstyle (2m-1)(2m+3)}F''_m
+\frac{\scriptstyle (m+2)(m+1)}{\scriptstyle (2m+3)(2m+5)} F''_{m+2} \right)
$$
$$
+ \frac{\gamma}3  \left(
-\frac{\scriptstyle m(m-1)(2m-5)}{\scriptstyle (2m-1)(2m-3)}F'_{m-2} +
\frac{\scriptstyle 8m(m+1)-4}{\scriptstyle (2m-1)(2m+3)}F'_m
+\frac{\scriptstyle (m+1)(m+2)(2m+7)}{\scriptstyle (2m+3)(2m+5)} F'_{m+2} \right)=0
$$
As $z\to \infty$, we neglect the terms proportional to  $\left(
\frac{F_n}{z}\right)$. The resulting equations set has the
solution
%$$
%F_1 = 3 F_0\left(1+O\left(\frac{1}{z}\right)\right)\,,\quad F_2= 5
%F_0 \left(1+O\left(\frac{1}{z}\right)\right)\,, ...,
%$$
\begin{equation}  \label{razloj}
 F_m = (2m+1) F_0 (z) \,
 % \left(1+O\left(\frac{1}{z}\right)\right)
\end{equation}
where $F_0(z)$ satisfies (\ref{Fmu}) for
$\overline{\mu}=\overline{\mu^2}=1$. (Actually, the coefficients
(\ref{razloj}) are the Legendre coefficients of the function
$2\delta(1-\mu)F_0(z;\alpha)$.) Combining (\ref{razloj}) with the
definition of ${\overline\mu(z)},{\overline{\mu^2}(z)}$ (\ref{opr}), we get
$$
{\overline\mu(z)} = 1 - O\left(\frac{1}{z}\right) \, \quad
{\overline{\mu^2}(z)} = 1 - O\left(\frac{1}{z}\right)
$$
The equation (\ref{w})  takes the form
$$
(1+\gamma) zw_{zz} + \left(\frac43 +2\gamma  -z\right) w_z +
\left(\frac{\alpha}{3} - \frac{4}{3(1+\gamma)} \right) w + 2\gamma
\left( \frac{1}{9z}  - \frac 1{(1+\gamma)} \right) w = 0
$$
This is equivalent to  Kummer degenerate hypergeometric equation
\cite{Kum}. The solutions of this equation are
\begin{equation} \label{reshw1}
w_1(z) = z^{-\frac23 \frac{\gamma}{1+\gamma}}
M\left(a,b;\frac{z}{1+\gamma}\right)\,
\end{equation}
$$
w_2(z)=z^{-1/3} M\left(1+a-b,2-b;\frac{z}{1+\gamma}\right) \,
$$
where $M$ is the Kummer function
$$
M(a,b,\zeta)=1+\frac ab \frac z{1!} + \frac{a(a+1)}{b(b+1)}
\frac{z^2}{2!}+... \,
$$
the parameters $a$ and $b$ are
$$
a=\frac{4-\alpha}3 \,  \quad b=\frac 23 \frac{2+\gamma}{1+\gamma}
$$
%$$ M_{1,2}(z)= M,U \left(\frac{4-\alpha}{3}, \frac{4+\gamma+\gamma
%\mu_1^2}{3(1+\gamma \mu_1^2)},\frac{\mu_0}{1+\gamma
%\mu_1^2}z\right)
%$$

We have found the general solution. Let us now check the boundary
conditions (\ref{granicay=0}), (\ref{granicax=0}). The solution
$w_2$ gives $F_0\sim z^{-1/3}$ as $z\to 0$. The correspondent
$\int  y^2 \frac{\partial f}{\partial y} d\mu $ does not vanish as
$y\to 0$. This contradicts to the boundary condition
(\ref{granicay=0}). Hence, the solution of our problem is $w_1$,
since it satisfies (\ref{granicay=0}).

The Kummer functions behave like $M(a,b,z)\sim e^{z}
z^{a-b}$ as $z\to \infty$ if $a$ is not negative  integer.
Therefore, for the corresponding values  $\alpha=4-3a$ and for
small $x$ we have $F(z) \sim z^{-\alpha/3}$, $fx^2 \sim
y^{-\alpha}$. This means that the no-flux condition on the
boundary $x=0$ (\ref{granicax=0}) is not satisfied.  Hence, to
satisfy both boundary conditions (\ref{granicay=0}) and
(\ref{granicax=0}) one should take the solution $w_1$  with the
values $\alpha$ that correspond to "discrete" spectrum
\footnote{$n=0$ is excluded since it does not satisfy the
normalization condition for $f$. }
\begin{equation}\label{alpha}
\alpha = 4 + 3 n\,,\qquad n= 1,2,3....
\end{equation}
For these values $\alpha$ the series $M$ contains a finite number
of terms, the leading term being $\sim z^n$.

The solution (\ref{reshw1}) together with  (\ref{alpha}) and
(\ref{razloj}) gives asymptotic behavior of the function $f$ as
$t\to \infty$. We stress that, according to (\ref{fF}),
(\ref{alpha}), the full probability density function $f(x,y,\mu)$
in the leading asymptotic term behaves like
$$
f(x,y,\mu) \sim x^{-9} F(z,\mu; 7)
$$
as $x\to \infty$. Below, we will need integrability of the
function $F$ only.

\section{Spatial distribution of vorticity. Singularity of
vorticity and pair correlation function. }
%Domains of growth and...

In the previous section we found an asymptotic solution for
probability distribution of $\bomega,\,\dot{\bomega}$. It is
important that the solution has power fall as $x\to\infty$. This
means a significant probability of large-amplitude fluctuations of
$|\bomega|$. This is the manifestation of intermittency in the
turbulence: in some spatial domains  the value $|\bomega|$ is much
larger than its average. The question is how the flow in this
domains should look like to provide the obtained asymptote.

For this purpose, let us define the probability density function for the
module of vorticity  based on the combined probability density of
$\bomega,\,\dot{\bomega}$ (\ref{sovmestxy}),(\ref{f}):
\begin{equation}\label{P}
P(x,t) =\int f(t,{\bf x},{\bf y}) x^2 d{\bf y} d\mu = \left<
\delta(x-|{\bf x}(t)|) \right>
\end{equation}
On the other hand, we independently define a probability density
$P_1(x,t)$ as space average of some realization of the turbulent
flow:
\begin{equation}\label{P1}
P_1(x,t) =  \frac{1}{V}\int \delta(x-X(t,{\bf r})) d{\bf r} .
\end{equation}
Here $X(t,{\bf r})= |\bomega(t,{\bf r})|$ is the vorticity module
at time $t$ and at the point ${\bf r}$, $V$ is the volume of space
occupied by the flow.

The first expression for the probability density $P(x,t)$ is the
ensemble average along a trajectory of liquid particle $\bxi(t)$,
and the second expression for $P_1(x,t)$ is the space average.

Owing to ergodicity (i.e. the equality of ensemble and  %under assumption that
space averages) we obtain
\begin{equation}\label{=}
P(x,t) = P_1(x,t)
\end{equation}

If the function  $P(x,t)$ is known, then it is possible to derive
space distribution of the vorticity module $X(t,{\bf r})$ using
(\ref{=}) and (\ref{P1}).  Since we are interested in possible
singularities and their surrounding, let us consider the limit $t\to\infty$
and next $x\to \infty$. 

Suppose that the singularity is reached at some surface.
Taking the point of origin on the surface and the axis $z$
perpendicular to it, we find
$$
P_1(x,t) = \frac 1V \int \delta\left( x-X(t,z)\right) d\sigma dz =
\left. \frac 1{|X'_z|} \right|_{X(t,z)=x}
$$
Here $d\sigma$ is the element of the surface area. The simplest
example of surface where the vorticity grows unrestrictedly large
is the tangential break of flow velocity. Indeed, in the case the
velocity is $V_0$ on one side of the contact surface and $-V_0$ on
the other side. Hence, the vorticity is concentrated on the
 surface.

Let us now consider the most interesting case: the maximum of
$X(t,{\bf r})$ is reached along a vortex line.  Then, choosing the
cylindrical variables $z,r,\phi$ with  $z$ axis oriented along the
line, we obtain
$$
P_1(x,t) = \frac 1V \int \delta\left( x-X(t,r)\right) r dr d\phi
dz = \left. \frac r{|X'_r|} \right|_{X(t,r)=x}
$$
In the case of point-like maximum, using spherical coordinates
$r,\theta, \phi$, we would obtain
$$
P_1(x,t) = \frac 1V \int \delta\left( x-X(t,r)\right) r^2 dr \sin
\theta d\theta d\phi  = \left. \frac {r^2}{|X'_r|}
\right|_{X(t,r)=x}
$$

Taking the limit $t\to \infty$,  with account of (\ref{=})  we find the
 spatial distribution of vorticity module in the
vicinity of singularity:
$$
X'(z) P(X) = 1 \, - \hbox{singular surface}
$$
\begin{equation}\label{x(r)}
X'(r_{\perp}) P(X) = r_{\perp} \, - \hbox{singular line}
\end{equation}
$$
X'(r) P(X) = r^2 \, - \hbox{singular point}
$$

In Section 4 we found the asymptotic expression for probability
density function $F_0=\int F(z,\mu,x)d\mu $ (\ref{reshw1}),
(\ref{alpha}). Integrating it with respect to $z$, we obtain the
function $P(x)$:
\begin{equation}\label{intF}
P(x)= \sum \limits_{\alpha} p_{\alpha}  x^{3-\alpha} \, \quad
\alpha=4+3n \, \quad n=1,2,3,...
\end{equation}
Combining  (\ref{intF}) with (\ref{x(r)}) and integrating
(\ref{x(r)}) in the vicinity of the singular point, we get in the
leading asymptotic term
\begin{equation}\label{nit}
X(z)\sim z^{-\frac13}\qquad  -\hbox{singularity of vortex
surface}
\end{equation}
\begin{equation}\label{2/3}
X(r_{\perp})\sim r_{\perp}^{-\frac{2}{3}}\qquad -\hbox{singularity
of vortex line}
\end{equation}
$$
X(r)\sim r^{-1}\qquad  -\hbox{singular point}
$$
We see that in the case of singular point, $X(r)$  diverges less
than any singularity
%in any solution of the
%Laplace equation (and this is just the Laplace equation that
%specifies
%(which is just responsible for
%all the point singularities of the vortex field).
specified by the Laplace equation. This means that the singularity of such kind
 cannot exist as the isolated point.

The most divergence in (\ref{2/3}) is given by the singularity
along the filament. We shall show that the dependence
 $X(r_{\perp})\sim r_{\perp}^{-\frac{2}{3}}$
corresponds to the Kolmogorov law. Consider the correlator of
transverse velocities 
\begin{equation}\label{K(r)}
K(r_{\perp})=\left<({\bf v}_{\perp}({\bf r},t) - {\bf v}_{\perp}(0,t))^2\right>
\end{equation}
Here ${\bf v}_{\perp}$ is the velocity component perpendicular to the
current line  $\bxi(t)$. According to the definition
(\ref{hdmoving}), the expression ${\bf v}_{\perp}({\bf r},t) -
{\bf v}_{\perp}(0,t)$ is identically equal to $\bomega \times {\bf
r}$ in the vicinity of the vorticity's singularity. Hence, as
$r\to 0$ we get
\begin{equation}\label{I}
K(r_{\perp})\propto \omega^2 r_{\perp}^2 \propto r_{\perp}^{2/3}
\end{equation}
Since the direction of the line is arbitrary, the turbulent
pulsations change it. From the isotropy it naturally
follows that the space average over the main scale of the
turbulence is equivalent to the average over the angles. The
expression (\ref{I}) then transforms to
\begin{equation}\label{II}
K(r)\propto r^{2/3}
\end{equation}

So, the main expression (\ref{2/3a}), (\ref{II}) follows naturally
from our consideration. It means that Kolmogorov's correlation
function (\ref{II})  is determined by the system of {\it vortex filaments}.
The input of regular part of the velocity to the correlation function $(\propto r^2)$ 
is negligible.

\section{Conclusion}

In this paper we investigated the small-scale structure generation 
in the developed  turbulence.  Here we summarize and discuss
briefly the main results.

1. Vortex structures

We considered the Navier-Stokes equation as   $\nu\to 0$. 
We derived the equations  describing the growth of
small-scale pulsations along the Lagrangian trajectory under the
action of the large-scale turbulence. We showed that the
small-scale part of vorticity grew exponentially in time. This
growth led to formation of a system of filaments and  surfaces  where
the vorticity grew intensively.  We derived the characteristic
parameters  of the vortex structure  growth in time.

 2. Singularity

We showed that in the non-dissipative limit $\nu\to 0$ the
absolute value of the vorticity $\omega$ tended to infinity along
the vortex filaments as $t\to\infty$. Notice that constructing the
probability density function  (\ref{f}) we linearized the
hydrodynamic equations near the Lagrangian trajectory
(\ref{omega}). We obtained the exponential growth  of vorticity
which  may be cut either by nonlinear corrections or  by viscosity
of the flow. Let us discuss now the nonlinear corrections.
The feedback effect of the small-scale pulsations could be estimated
by comparison of the energy density of small-scale pulsations with the energy
of large-scale ones.
The energy density of
the main pulsations on the scale $L$ is
$$
E_0=\frac{1}{2}\rho U^2 \,
$$
here $U$ is the velocity of the large-scale pulsations. From
(\ref{nit}) we estimate  the velocity  of
the filament having width $r_0$: $ v_n\sim U \left( r_0/L\right)^{1/3} $.
Taking into account the part of volume occupied by the filaments, we
obtain the  relation between the energy densities:
$$
\frac{E_n}{E_0} \sim N_f  \left( \frac{v_n}{U}\right) ^2 \cdot
\left( \frac{r_0}{L} \right)^2 \sim N_f \left(  \frac{r_0}{L}
\right)^{8/3} \ll 1 .
$$
Here $N_f$ is the ratio of the number of filaments to the number of
large-scale vortex in the volume unit.

Thus, we see that the feedback effect of small-scale pulsations on the large-scale
ones is insignificant.
Hence, the singularity should be cut off by
viscosity. The situation is quite analogous to that in supersonic
hydrodynamics: the singularities (strong and weak discontinuities) in Euler
flow are cut off by viscosity.

3. Correlation function

We found the solution of the equation describing  the vorticity
distribution in the vicinity of the vortex filaments.
The solution had the form $|v_{\perp}|\propto r_{\perp}^{1/3}$ in
the plane perpendicular to the vortex filament.  This was the solution
that
determined the form of the pair correlation function in small
scales. Thus, we found the velocity correlation function 
(\ref{2/3a}), (\ref{II}) in the steady-state turbulent flow directly
from the Navier-Stokes equation in non-viscous limit $\nu\to 0$.

4. Intermittency

According to the   Kolmogorov-Richardson assumption the energy flux
in turbulent flow cascades from larger scales to smaller ones and
dissipates at the smallest scales uniformly  in space and time.
Landau pointed out that this assumption was controversial (see \cite{Fr}
\S 6.4). Gurvich in 1960  \cite{Gu}, and later the other researchers
discovered experimentally a very strong time and space  inhomogeneity of
velocity and energy flux. This property of turbulence is called
intermittency. Variety of approaches to this  effect was considered
by many authors (see monographs \cite{MY},\cite{Fr} and citations
therein).

Let us list the intermittency features that follow from
the  presented theory.

1) The vorticity  distribution in space is very inhomogeneous.
Near the vortex axis it could possess the value many times exceeding
its average.

2) Even moments of the correlation functions should grow with
number of the moment.

3) The energy dissipation in the developed turbulent flow is
localized near the axes of vortex filaments and the vortex surfaces. It
is distributed very inhomogeneously in space and  time due to  the
vortex structures motion. Besides, the strong nonuniform
dissipation is the most pronounced manifestation of the intermittency
\cite{MS}.

Note that the filaments give the maximum degree of
singularity as $t\to\infty$ and are responsible for the form of the
pair correlation function. However, the surface-type singularities
may affect the dissipation process, since  they could occupy a
significant part of the volume of the flow.

\vspace{0.5cm}

The authors are grateful to V.L. Ginzburg for the attention  to this work, and to
A.S. Gurvich, V.S. Lvov, E.A. Kuznetsov, S.M. Apenko, V.V. Losyakov and M.O. Ptitsyn for 
useful discussions. 

This research was partially supported by the RAS Presidium Program 
"Mathematical methods in nonlinear dynamics".

\vspace{1cm}

{\bf Appendix}

Let us consider an axially symmetric flow. The hydrodynamic equations 
in the cylindric coordinates take the form
$$
\frac{\partial v_r}{\partial t} + v_r \frac{\partial v_r}{\partial r} +
v_z \frac{\partial v_r}{\partial z} -\frac{v_{\phi}^2}{r} = -\frac{\partial p}{\partial r} \qquad
{\mbox \hfil (A.1)}
$$
$$
\frac{\partial v_{\phi}}{\partial t} + v_r \frac{\partial v_{\phi}}{\partial r} +
v_z \frac{\partial v_{\phi}}{\partial z} +\frac{v_{\phi}v_r}{r} = 0
\quad\qquad {\hfill (A.2)}
$$
$$
\frac{\partial v_z}{\partial t} + v_r \frac{\partial v_z}{\partial r} +
v_z \frac{\partial v_z}{\partial z} = -\frac{\partial p}{\partial z}
\quad\quad\quad\qquad {\mbox  (A.3) }
$$
$$
\frac{1}{r}\frac{\partial}{\partial r}(r v_r) + \frac{\partial v_z}{\partial z} =0\qquad
\qquad\qquad\quad\quad \, (A.4)
$$
Here  $v_r$, $v_{\phi}$, $v_z$ are the  radial, asimutal   and parallel to the cylinder's axis 
velocity components, respectively.

We search  a solution of the system  (A.1) -- (A.4) in the form
$$
v_{\phi} = \omega r\,, \quad v_r = a r\,,\quad v_z = b z \qquad \quad \, (A.5)
$$
Then the pressure should be 
$$
p(r,z,t) = \frac{P_1(t)}{2} r^2 + \frac{P_2(t)}{2} z^2
$$
From (A.4) follows a relation between   $a$ and  $b$:
$$
2a + b = 0\,,\qquad \qquad  \qquad \qquad (A.6)
$$
This relation expresses the volume conservation in the liquid. Indeed, let us consider a cylindric drop
with radius $R(t)$ and length $Z(t)$. Then from (A.5) follows
$$
\dot{R} = a(t) R\,,\quad \dot{Z} = b(t)Z\,,
$$
$$
\hbox{hence}
\qquad R(t) = R_0\,exp\left(\int_0^t a(t_1) dt_1
 \right)\,,\quad  Z(t) = Z_0\,exp\left(\int_0^t b(t_1) dt_1 \right)
$$
The cylinder volume at arbitrary time $t$ is 
$$
\pi R(t)^2 Z(t) = \pi R_0^2Z_0 \exp\int_0^t\left(2a(t_1) + b(t_1)\right) dt_1 = \pi R_0^2 Z_0.
$$
We see that volume conserves.  For example, if $b>0$ then the cylinder  stretches, and its
transversal radius decreases.  

Combining   (A.5) with (A.1) -- (A.4), we obtain a system of ordinary differential equations: 
$$
\dot{a} + a^2 - \omega^2 = - P_1   \qquad\qquad\qquad
$$
$$
\dot{\omega} + 2 a \omega = 0   \qquad\qquad\qquad  \quad (A.7)
$$
$$
\dot{b} + b^2 = - P_2 \qquad\qquad\qquad
$$
Note that the system allows one arbitrary function of time.  Actually, with account of (A.6)
we have four equations and five unknown functions: $a\,, b\,, \omega\,, P_1\,, P_2$. 
Without loss of generality one can choose  $P_2(t)$ as such arbitrary function.
We also note that the change of vorticity $\omega(t)$ is connected unambiguously with  change
of the "cylinder length" $Z(t)$: $\omega(t) = \omega_0 Z(t)/Z_0$.

Differentiating the second equation of the system (A.7) and substituting other equations, we get
$$
\ddot{\omega} = -P_2(t)\, \omega
$$
This equation is a particular case of  (\ref{omega}).
We assume that $P_2(t)$ is rather complicated "random" function and its time average
is zero.  Then the time intervals when  $P_2(t)>0$ and $P_2(t)<0$ are  equally probable.
However, at $P_2(t)>0$ the function $\omega(t)$ oscillates, the oscillation amplitude changing
weakly. To the contrary, at  $P_2(t)<0$ the function $\omega(t)$ grows exponentially.
It is clear that in the average the value  $\omega$ grows. Since 
$\omega$ and $Z$ are proportional, such growth means "stretching out" the cylinder.


\newpage

\begin{thebibliography}{40}

\bibitem{LL} L.D.Landau, E.M.Lifshitz "Hydrodynamics" Pergamon Press 1975, ch.3.

\bibitem{MY} A.S,,Monin A.M.Yaglom "Statistical Fluid Mechanics" vol.1 Ed.J.Lumley,
MIT Press, Cambridge, MA, 1971; vol.2  1975.

\bibitem{Fr} U.Frisch, "Turbulence. The Legacy of A.N.Kolmogorov" Cambridge University
Press, 1995

\bibitem{Kol} A.N.Kolmogorov,  Doklady Academy of Science USSR, 
30, 9 - 13, 1941; 31, 583 - 540, 1941; 32, 16 - 18, 1941 (in Russian)

\bibitem{OB} A.M.Obukhov,  Doklady Academy of Science USSR,
32, 22-24, 1941;  (in Russian)

\bibitem{ZLF} V.E.Zakharov, V.S.L'vov, G.Falkovich, "Kolmogorow spectra of turbulence", Springer,
Berlin, 1992

\bibitem{Vis} M.J.Vishik, A.F.Fursikov "Mathematical problems of statistical hydrodynamics"
Kluwer, Dordrecht, 1988

\bibitem{Avis} C.Foias, O.Manley, R.Rosa and R.Temam "Navier-Stokes equations and Turbulence"
Cambridge Univ. Press 2001

\bibitem{Saff} P.G.Saffman "Vortex dynamics" Cambridge Univ.Press, Cambridge, 1992

\bibitem{L'v} V.S.L'vov, I.Procaccia "Analytic calculation of anomalous exponents in turbulence:
Using the fusion rules to flush out a small parameter" Phys.Rev.E, v.62, N 6, 8037 - 8057, (2000).

\bibitem{Yah} V.Yakhot "Probability density in strong turbulence" arXiv:physics/0512102 v3 (2005).

\bibitem{LLag} V.I.Belincher, V.S.L'vov, Sov.Phys. JETP 66, 349 (1977).

\bibitem{Kuz} E.A.Kuznetsov and V.P.Ruban, JETP, 91, 775-785 (2000).

\bibitem{NWL} A.Noullez, G.Wallace, W.Lempert, R.Miles, U.Frisch 
J.Fluid. Mech., 339; 287 - 307, (1997)

\bibitem{KL} V.I.Klyatskin "Dynamics of Stochastic Systems" Fizmatlit, 2003

\bibitem{Kum} M.Abramowitz  I.Stegun "Handbook of Mathematical Functions" National
Bureau of Standards, 1964

\bibitem{Gur1971} A.S.Gurvich, V.V.Pachomov, A.M.Cheremuchin,  Radiofizika, v.7,
76-80, (1971).

\bibitem{Gu} A.S.Gurvich, Izvestiya Academy of Sci USSR, geofizika, 7, 1042-1055, 1960

\bibitem{MS}C.M.Menevean, K.R.Sreenivasan, "The multifractal nature of turbulent energy
dissipation", J.Fluid Mech. 224, 429 - 484, 1991



\end{thebibliography}
\end{document}